\documentclass[reprint,showpacs,showkeys,preprintnumbers,amsmath,amssymb,aip,apl]{revtex4-1}

\usepackage{graphicx}
\usepackage{float}
\usepackage{color}

\begin{document}

\preprint{Rong Shan et al., PdLuBi}

\title{Electronic and crystalline structures of {\it zero band-gap} PdLuBi thin films grown epitaxially on MgO(100)}

\author{Rong Shan}
\affiliation{Institut f{\"u}r Anorganische und Analytische Chemie,
             Johannes Gutenberg - Universit{\"a}t, 55099 Mainz, Germany.}
\affiliation{IBM Almaden Research Center
             San Jose, CA 95120, USA.}
             
\author{Siham Ouardi}
\affiliation{Max Planck Institute for Chemical Physics of Solids,
             01187 Dresden, Germany.}
            
\author{Gerhard~H. Fecher}
\affiliation{Max Planck Institute for Chemical Physics of Solids,
             01187 Dresden, Germany.}

\author{Li Gao}
\affiliation{IBM Almaden Research Center
             San Jose, CA 95120, USA.}
             
\author{Andrew Kellock}
\affiliation{IBM Almaden Research Center
             San Jose, CA 95120, USA.}

\author{Kevin~P. Roche}
\affiliation{IBM Almaden Research Center
             San Jose, CA 95120, USA.}

\author{Mahesh~G. Samant}
\affiliation{IBM Almaden Research Center
             San Jose, CA 95120, USA.}
                         
\author{Carlos~E. Vidal Barbosa}
\affiliation{Max Planck Institute for Chemical Physics of Solids,
             01187 Dresden, Germany.}      

\author{Eiji Ikenaga}
\affiliation{Japan Synchrotron Radiation Research Institute (JASRI), SPring-8, Hyogo 679-5198, Japan.}

\author{Claudia Felser}
%\email{felser@cpfs.mpg.de}
\affiliation{Max Planck Institute for Chemical Physics of Solids,
             01187 Dresden, Germany.}

\author{Stuart~S.~P. Parkin}
\email{Stuart.Parkin@us.ibm.com}
\affiliation{IBM Almaden Research Center
             San Jose, CA 95120, USA.}

\date{\today}

\begin{abstract}

Thin films of the proposed topological insulator PdLuBi - a Heusler compound with the $C1_b$ 
structure - were prepared on Ta-Mo-buffered MgO(100) substrates by co-sputtering 
from PdBi$_2$ and Lu targets. Epitaxial growth of high-quality 
PdLuBi films was confirmed by X-ray spectrometry and reflection high-energy 
electron diffraction. The root-mean-square roughness of the films was as low as 
1.45~nm, although the films were deposited at high temperature. The film 
composition is close to the ideal stoichiometric ratio. The valence band spectra 
of the PdLuBi films, observed by hard X-ray photoelectron spectroscopy, 
correspond perfectly to the ab-initio-calculated density of states.

\end{abstract}



\maketitle

%\section{Introduction} %%%%%%%%%%%%%%%%%%%%%%%%%%%%%%%%%%%%%%%%%

Zero energy consumption in electron transport, 
spin channel splitting, protection of spins from thermal activity, and other interesting, 
recently proposed, physical properties in topological insulators (TIs) would greatly improve the performance 
of semiconductor devices, achieve spin control in spintronics devices without 
application of magnetic fields, and provide an excellent platform for the 
application of quantum physics, e.g., quantum 
computers~\cite{KMe05,BHZ06,QHZ08,KWB07}. Until now, most 
studies of TIs have focused on HgTe/CdTe/HgTe quantum wells 
with two-dimensional (2D) topological states~\cite{KWB07} and three-dimensional (3D) 
TIs, Bi$_{1-x}$Sb$_x$, Bi$_{1-x}$Te$_x$, and
Sb$_{1-x}$Te$_x$~\cite{FKM07,ZLQ09,CAC09}. Electrical properties 
related to the quantum spin Hall effect have only been observed in HgTe/CdTe/HgTe 
quantum wells, which have been fabricated by molecular beam epitaxy and used as 
rudimentary devices. Even for HgTe/CdTe/HgTe quantum wells, the difficult and  
expensive fabrication of the devices has impeded further study. In order to resolve
this problem, we suggest a new platform based on half-Heusler compounds for the study of TIs~\cite{CQK10}.
A band gap arises in half-Heusler compounds when the total number of valence 
electrons equals 18 as a result of a closed-shell electronic configuration~\cite{JKW00}, 
although the material consists of three metals. The $\Gamma_6$ and $\Gamma_8$ 
bands are inverted in some semiconducting Heusler compounds with heavy elements 
as a result of strong spin--orbit coupling~\cite{CQK10,XYF10}. This leads to a TI state 
with a zero band gap that can be used in 2D TI quantum well structures. 
Considering the large range of lattice constants corresponding to the variety of 
materials and the easy fabrication conditions of Heusler compounds, we 
suggest that TI devices with Heusler compounds may become 
feasible for applications. For this purpose, high-quality epitaxial PdLuBi films were fabricated by co-sputtering.

%\section{Experimental details} %%%%%%%%%%%%%%%%%%%%%%%%%%%%%%%%%%%%%%%%%

Samples with the structures MgO(100)/Mo~(2~nm)/Ta~(20~nm)/Mo~(1~nm)/ PdLuBi~(40~nm)/MgO~(17~nm)
were prepared by direct-current magnetron co-sputtering under 
an Ar atmosphere. The base pressure of the sputtering chamber was below $5\times10^{-7}$~Pa. 
The angle of inclination between each target and the substrate was about 
$9^\circ$ and the distance between the sources and target was about 7.95 inches. The 
substrate was rotated during film growth. The composition of the PdLuBi thin film was analyzed by Rutherford 
backscattering spectrometry and particle-induced X-ray emission. 
The crystalline structure was examined by X-ray diffraction (D8-Discover, Bruker AXS Inc.). 
The surface morphologies of the films were analyzed in situ using reflection high-
energy electron diffraction (RHEED, STAIB Instrument Inc.). Ex situ surface 
analysis was carried out using atomic force microscopy (5600LS, Agilent 
Technologies Inc.). The valence band spectra of the PdLuBi films were observed using 
hard-X-Ray photoelectron spectroscopy (HAXPES) at BL47XU of Spring-8 (Japan). 
For details of the HAXPES experiments, see References~\cite{FBG08,OFK11,KSF11}.

Mo and Ta layers were deposited at $200^\circ$C, and then annealed at 
$900^\circ$C. The Mo layer was used to induce growth of Ta(100) on MgO(100) 
instead of Ta(110). After this process, a single-crystalline film of Ta--Mo(100) 
was built on MgO(100) with a $45^\circ$ in-plane rotation. Because Mo and Ta can 
be miscible with each other at high temperatures, the lattice constant of Ta--Mo can 
be tuned by precise adjustment of the composition. In this study, the lattice 
constant of PdLuBi was slightly smaller than twice the lattice constant of Ta.  A 
thin Mo layer was therefore used between the Ta and PdLuBi layers. The 
root-mean-square roughness of the Ta--Mo layer was below 3~{\AA}. The PdLuBi layer was 
deposited at $800^\circ$C.  To obtain a semiconducting Heusler film of high 
quality, a high substrate temperature is usually necessary. However, as a result of the 
low bonding energy between Bi atoms, Bi almost disappears above a substrate temperature of
$600^\circ$C when the film is 
sputtered from three separate targets. On the 
other hand, the melting point of PdBi$_2$ ($\approx 480^\circ$C) is much higher 
than that of Bi ($\approx 271^\circ$C), implying a higher bonding energy of Pd--Bi.
Bi was therefore sputtered from a PdBi$_2$ target to reduce the problem of Bi loss at high growth temperatures. 
In this case, the composition ratio of Pd:Bi could be tuned using the annealing 
temperature and the composition ratio of Lu:Pd could be tuned using the sputtering 
power. Using this procedure, films with a composition ratio of Pd:Lu:Bi = 33.5:33.4:33.1 
(at\%; $\pm1$\%) were obtained at $800^\circ$C.

\begin{figure}
\centering
   \includegraphics[width=6cm]{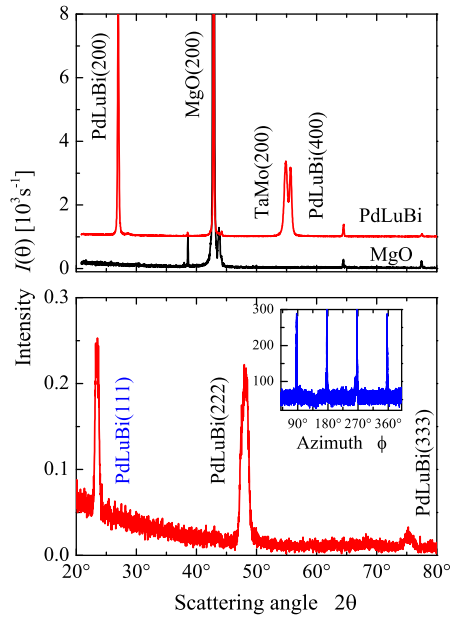}
   \caption{(Color online) X-ray diffraction patterns of thin PdLuBi films on Ta--Mo-buffered MgO(100).\\
            (a) (100) polar scan of PdLuBi (red curve); 
                the black solid curve indicates the background from the MgO(100) substrate. 
            (b) (111) polar scan of PdLuBi and azimuth scan of PdLuBi(111) (inset).}
\label{fig:xrd}
\end{figure}

Figure~\ref{fig:xrd} shows the X-ray diffraction pattern of a PdLuBi thin film 
on Ta--Mo-buffered MgO(100). Only the (200) and (400) diffraction peaks of the PdLuBi film are 
observed in the (100) polar $2\theta$-scan (see \ref{fig:xrd}(a)). The lattice 
constant of the PdLuBi film is $\approx 6.56$~{\AA} and corresponds to that 
reported for the bulk material~\cite{HSR02}. The (111) polar scan and the azimuth 
$\phi$-scan of the PdLuBi(111) diffraction peak (inset of \ref{fig:xrd}(b)) with 
four-fold symmetry prove epitaxial growth of the PdLuBi film. 

% Figure 1 %%%%%%%%%%%%%%%%%%%%%%%%%%%%%%%%%%%%%%%%%%%%%%%%%%%%%%%%%
%%%%%%%%%%%%%%%%%%%%%%%%%%%%%%%%%%%%%%%%%%%%%%%%%%%%%%%%%%%%%%%%%%%%

\begin{figure}
\centering
   \includegraphics[width=4cm]{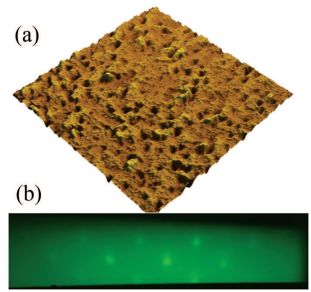}
   \caption{(Color online) Morphology of thin PdLuBi films. \\ 
           (a) Atomic force microscopy three-dimensional surface representation of PdLuBi with cap layer, 
           (b) reflection high-energy electron diffraction pattern of PdLuBi.}
\label{fig:morph}
\end{figure}

The surface morphology of the film is shown in Figure~\ref{fig:morph}(a). The 
root-mean-square roughness is around 1.45~nm, although the film was deposited 
at $800^\circ$C. The RHEED pattern of the PdLuBi film without a cap layer is shown in Figure~\ref{fig:morph}(b). 
It is known that single-crystalline films with an atomically flat 2D surface produce a stripe 
pattern, whereas single-crystalline films with an atomically rough (3D) surface 
produce a transmission pattern, i.e., a set of broad spots~\cite{ICO04}. 
Polycrystalline films and textured films without in-plane orientation  
form a set of concentric circles instead. It is clear from Figure~\ref{fig:morph} that a 
spotty pattern was observed. This indicates that a single-crystalline PdLuBi 
film with an atomically rough surface---either via a 3D island (Volmer--Weber) growth 
mode or a layer-plus-island (Stranski--Krastanov) growth mode---was obtained 
successfully.

% Figure 2 %%%%%%%%%%%%%%%%%%%%%%%%%%%%%%%%%%%%%%%%%%%%%%%%%%%%%%%%%
%%%%%%%%%%%%%%%%%%%%%%%%%%%%%%%%%%%%%%%%%%%%%%%%%%%%%%%%%%%%%%%%%%%%

%\section{Results and discussion} %%%%%%%%%%%%%%%%%%%%%%%%%%%%%%%%%%%%%%%%%

\begin{figure}
\centering
   \includegraphics[width=8.5cm]{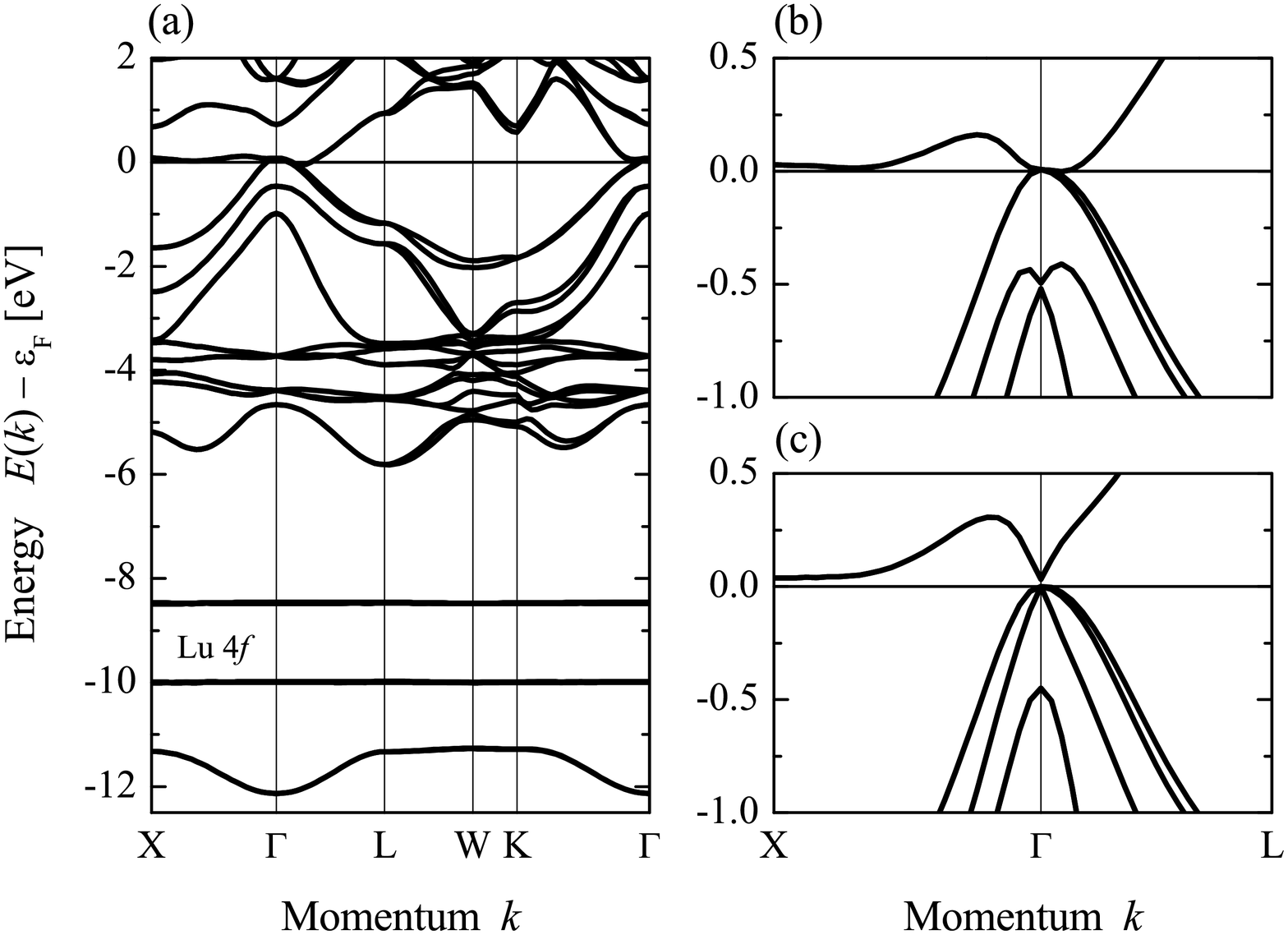}
   \caption{Electronic structure of PdLuBi.\\
            (a) shows the complete band structure for the relaxed lattice parameter. 
            (b) and (c) show the band structure in the vicinity of the Fermi energy for the 
            thin film lattice parameter and a 5\% compressed lattice parameter, respectively.
            In (c), a small band gap of width 40~meV is opened.
            Calculations were performed with spin--orbit interactions for all elements and LDA$+U$
            for the Lu $4f$ electrons.}
\label{fig:es}
\end{figure}

The electronic structure of PdLuBi was calculated using {\sc Wien2k}~\cite{BSM01}. 
More details are given in Reference~\cite{OFB10}. Spin--orbit 
interactions were included for all atoms and LDA$+U$ with the self-
interaction correction double-counting scheme for Lu. An effective Coulomb 
energy of $U_{eff}=U-J=0.5$~Ry was used to reproduce the measured energy of the 
Lu $4f$ doublet; $a=6.6787$~{\AA} was found for the relaxed lattice parameter. 
The result of the electronic structure calculations is shown in 
Figure~\ref{fig:es}. The compound turns out to be similar to a {\it zero} band 
gap semiconductor (compare~\cite{OSF11,SOF12} and references there). The optical 
gap at $\Gamma$ is closed. The valence and conduction bands have a small overlap 
of about 130~meV. A fraction of about $10^{-3}$ electrons is found in the 
conduction band. This situation may change for slightly smaller lattice 
parameters (see Figure~\ref{fig:es}(c)) or a slight tetragonal distortion. Both effects are principally able 
to open up the band gap again at $\Gamma$ (compare also Reference~\cite{XYF10}). A slight tetragonal distortion may 
particularly occur in thin films with pseudomorphic growth.

% Figure 3 %%%%%%%%%%%%%%%%%%%%%%%%%%%%%%%%%%%%%%%%%%%%%%%%%%%%%%%%%
%%%%%%%%%%%%%%%%%%%%%%%%%%%%%%%%%%%%%%%%%%%%%%%%%%%%%%%%%%%%%%%%%%%%

\begin{figure}
\centering
   \includegraphics[width=7.5cm]{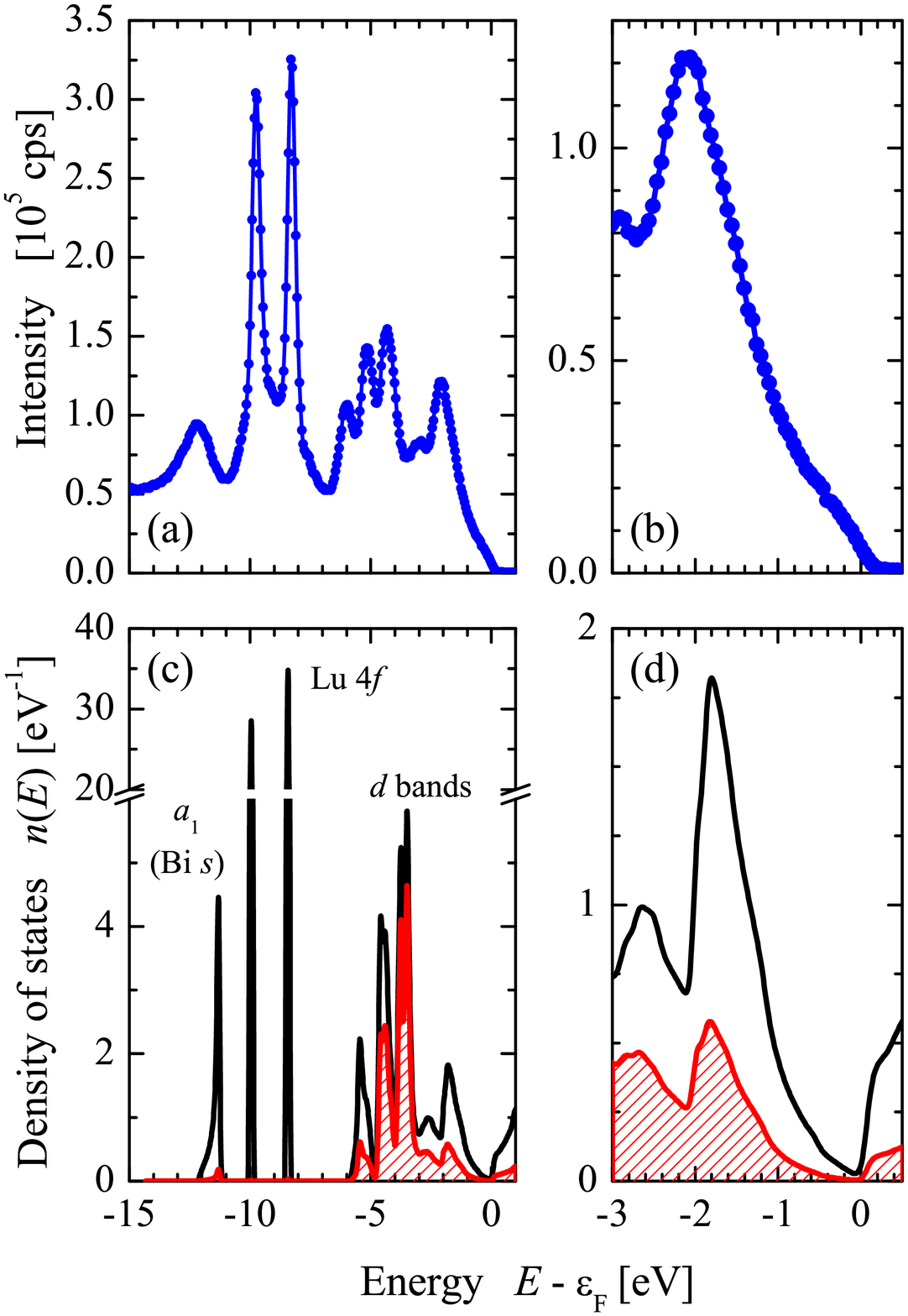}
   \caption{(Color Online) Valence band of PdLuBi.\\
            (a)--(b) show the valence band spectra 
            taken with a photon energy of about 8~keV.
            (c)--(d) show the density of states, and the density localized at the Pd atoms
            is marked by the shaded area.
            (b) and (d) show the valence band close to the Fermi energy on an enlarged scale.}
\label{fig:haxpes}
\end{figure}

Valence band spectra of PdLuBi are presented 
in Figure~\ref{fig:haxpes} and compared to the calculated density of states. 
Figure~\ref{fig:haxpes}(a) shows the valence band spectra excited by photons of 
energy about 8~keV. The high intensity with a maximum between -12.5 and -11~eV 
corresponds to excitation of the low-lying Bi $s$ states with $a_1$ symmetry, as is seen by 
comparison with the density of states. The Lu $4f_{5/2}$-$4f_{7/2}$ spin--orbit 
doublet is found at energies of $E_{5/2}= -9.76$~eV and $E_{7/2}= -8.31$~eV, with a
spin--orbit splitting of $\Delta_{\rm SO}=1.45$~eV. The Lu $4f$ states reside on top 
of the states from the MgO protective overlayer that usually appear at about -9~eV. 
Such MgO states were previously also observed in the spectra from MgO-covered 
Co$_2$MnSi~\cite{FBG08}. The upper part of the valence spectra above -6.5~eV 
exhibits the typical structure of the PdLuBi valence $d$ bands with four major 
maxima (-6, -5.16, -4.31, -2.1~eV) and a smaller one at -2.9~eV. The splitting 
of the states at about -4~eV is not resolved in the spectrum. Slight energy 
differences between the maxima in the density of states and maxima of the 
spectrum are observed. These energy shifts are larger for states further away from 
the Fermi energy (for example 0.3~eV at the -2.1~eV maximum and 0.6~eV at the 
-5.16~eV maximum). This is a typical effect of the photoemission process and 
emerges from the complex self-energy of the photoelectrons interacting with the 
remaining $N-1$ electron system. The spectra exhibit no clear cut-off at the 
Fermi energy as is observed when a metallic-type density is terminated by the 
Fermi--Dirac distribution, but a rather smooth, linearly decreasing intensity is 
observed. A similar behavior was found in bulk materials of the {\it zero} band 
gap Heusler compounds PtYSb~\cite{OSF11} and PtLuSb~\cite{SOF12}. A splitting
of the La $3d$ states was reported for thin PtLaBi films grown on YAlO$_3$(001) by three-source
magnetron co-sputtering~\cite{MSF12}. The appearance of the satellite peaks is explained as being the result of
charge transfer, as appears, for example, in oxides. The PdLuBi films reported here did not exhibit any 
additional splitting of core levels, which also indicates their high quality with respect to impurities, structure, and composition.

% Figure 2 %%%%%%%%%%%%%%%%%%%%%%%%%%%%%%%%%%%%%%%%%%%%%%%%%%%%%%%%%
%%%%%%%%%%%%%%%%%%%%%%%%%%%%%%%%%%%%%%%%%%%%%%%%%%%%%%%%%%%%%%%%%%%%

%\section{Summary and Conclusions} %%%%%%%%%%%%%%%%%%%%%%%%%%%%%%%%%%%%

In summary, high-quality epitaxial thin films of the Heusler TI PdLuBi were successfully prepared. 
The composition of Pd:Lu:Bi in the film grown at $800^\circ$C is stoichiometric, with ratios of 1:1:1. 
The valence band spectra of thin PdLuBi films observed by HAXPES perfectly match 
the density of states from first principles calculations. PdLuBi has a 
similar band structure to that of HgTe and thus may be used for the construction of 
2D TI quantum well structures.

%Acknowledgement %%%%%%%%%%%%%%%%%%%%%%%%%%%%%%%%%%%%%%%%%%%%%%%%%%%%%
%\bigskip
\begin{acknowledgments}
Financial support by the DFG-JST (projects P~1.3-A and P~2.3-A in research unit FOR 1464 {\it 
ASPIMATT}) is gratefully acknowledged. HAXPES was performed at BL47XU of SPring-8 
with approval of JASRI (Proposal No.~2012A0043).
\end{acknowledgments}

%\bibliography{PdLuBi}

\end{document}